%% file: lcws21.tex
\pdfoutput=1
\documentclass[12pt]{article}
\usepackage{graphicx}
\usepackage{subcaption}
\usepackage{amssymb}
\usepackage{amsmath}
\usepackage{amsfonts}
\usepackage{multirow}

\usepackage{xcolor}
\usepackage{url}
\usepackage{ulem}
\usepackage{float}
\usepackage{float}

\usepackage{pifont}

\usepackage{soul}
\usepackage[english]{babel}
\usepackage[T1]{fontenc}
\usepackage{lmodern}

\usepackage{dcolumn}
\usepackage[shortlabels]{enumitem}

\usepackage[utf8]{inputenc}  
\usepackage{bbm}

\usepackage{hyperref}
\hypersetup{
  colorlinks   = true, 
  urlcolor     = blue, 
  linkcolor    = black, 
  citecolor   = black 
}

\usepackage[numbers,sort&compress]{natbib}
\graphicspath{{plots/}}

\input{paperdef.tex}

\evensidemargin \oddsidemargin
\renewcommand{\arraystretch}{1.2}

\newcommand{\fnalbnlsig}{\htrs{0.8}}   
\newcommand{\newcen}{\htrs{206.1}}     
\newcommand{\newunc}{\htrs{4.1}}       
\newcommand{\newdiff}{\htrs{25.1}}    
\newcommand{\newdiffunc}{\htrs{5.9}}   
\newcommand{\newdiffsig}{\htrs{4.2}}   
\newcommand{\htrs}[1]{{\color{black} #1}}

\newpage

\allowdisplaybreaks
\begin{document}
\thispagestyle{empty}

\def\thefootnote{\fnsymbol{footnote}}

\begin{flushright}
\mbox{}
IFT--UAM/CSIC--21-053
\end{flushright}

\vspace{0.3cm}

\begin{center}

{\large\sc 
{\bf 
Improved \boldmath{$(g-2)_\mu$} Measurements and Supersymmetry:\\[.5em]
Implications for \boldmath{$e^+e^-$} colliders
}}

\vspace*{.7cm}

{\sc
Manimala Chakraborti$^{1}$%
\footnote{Talk presented at the International Workshop on
Future Linear Colliders (LCWS2021), 15-18 March 2021. C21-03-15.1.}%
, Sven Heinemeyer$^{2,3,4}$%
~and Ipsita Saha$^{5}$%
}

\vspace*{.5cm}

{\sl
$^1${Astrocent, Nicolaus Copernicus Astronomical Center of the Polish
Academy of Sciences, ul.\ Rektorska 4, 00-614 Warsaw, Poland\\}

\vspace*{0.1cm}

$^2$Instituto de F\'isica Te\'orica (UAM/CSIC), 
Universidad Aut\'onoma de Madrid, \\ 
Cantoblanco, 28049, Madrid, Spain

\vspace*{0.1cm}

$^3$Campus of International Excellence UAM+CSIC, 
Cantoblanco, 28049, Madrid, Spain 

\vspace*{0.1cm}

$^4$Instituto de F\'isica de Cantabria (CSIC-UC), 
39005, Santander, Spain
\vspace*{0.1cm}

$^5$Kavli IPMU (WPI), UTIAS, University of Tokyo, Kashiwa, Chiba 277-8583, Japan
}  

\end{center}

\vspace*{0.1cm}

\begin{abstract}
\noindent
The persistent $3-4\,\sig$ discrepancy
between the experimental result from BNL
for the anomalous magnetic moment of the
muon, \gmin2, and its Standard Model (SM) prediction,
was confirmed recently by the ``MUON G-2'' result from Fermilab.
The combination of the two measurements yields a deviation of
$\newdiffsig\,\sig$ 
from the SM value.
Here, we review an analysis of the parameter space of the
electroweak (EW) sector of the Minimal Supersymmetric Standard Model
(MSSM), which can provide a suitable explanation of the
anomaly while being in full agreement with other latest experimental
data like the 
direct searches for EW particles at the LHC and dark matter (DM)
relic density and direct detection constraints.
Taking the lightest supersymmetric particle (LSP) (the lightest
neutralino in our case) 
to be the DM candidate, we discuss the case of a mixed bino/wino LSP, which
can account for the full DM relic density of the universe and that of
wino and higgsino DM, where we take the relic density only as an upper bound.
We observe that an upper limit of $\sim 600 \gev$ can be obtained for
the LSP and next-to (N)LSP masses establishing clear search targets for
the future HL-LHC EW searches, but in particular for future high-energy
$e^+e^-$~colliders, such as the ILC or CLIC. 
\end{abstract}


\def\thefootnote{\arabic{footnote}}
\setcounter{page}{0}
\setcounter{footnote}{0}

\newpage


\section{Introduction}
\label{sec:intro}
While the LHC is yet to find any sign of new physics, indirect searches
such as, low-energy experiments, astrophysical measurements etc.\
are providing crucial hints of beyond the SM (BSM) scenarios. Specifically,
the sustained deviation of $3$-$4\,\sigma$ in the anomalous magnetic
moment of muon, \gmin2, 
between the theoretical prediction of the SM~\cite{Aoyama:2020ynm}
(see \citere{CHS3} for a full list of references) and
the experimental observation by the Brookhaven National Laboratory
(BNL)~\cite{Bennett:2006fi} has long been standing in support of new 
physics scenario. The BNL measurement,
\begin{align}
\amu^{\rm BNL} &= 11 659 209.1 (5.4) (3.3) \times 10^{-10}~,
\label{gmt-BNL}
\end{align}
when compared with the SM prediction
\begin{align}
\amu^{\rm SM} &= (11 659 181.0 \pm 4.3) \times 10^{-10}~,
\label{gmt-sm}
\end{align}
leads to a deviation of
\begin{align}
\Delta\amu^{\rm old} &= (28.1 \pm 7.6) \times 10^{-10}~,
\label{gmtdiff-BNL}
\end{align}
corresponding to a $\sim 3.7\,\sig$ discrepancy. The new result from Fermilab
``MUON G-2'' collaboration~\cite{Grange:2015fou}
was announced recently~\cite{Abi:2021gix},
which is within
$\fnalbnlsig\,\sig$ in agreement with  the older BNL result on \gmin2.
The combination of the two results was given as
\begin{align}
\amu^{\rm exp} = (11 659 \newcen \pm \newunc) \times 10^{-10},
\label{gmt-comb}
\end{align}
yielding a new deviation from the SM prediction of
\begin{align}
\De\amu^{\rm new} = (\newdiff \pm \newdiffunc) \times 10^{-10},
\label{gmtdiff-new}
\end{align}
corresponding to a discrepancy of $\newdiffsig\,\sig$.

This result particularly upholds one of the
leading candidates of BSM theories, the Minimal Supersymmetric Standard Model  
(MSSM)~\cite{Ni1984,Ba1988,HaK85,GuH86}.
The deviation in \refeq{gmtdiff-new} can easily be explained
in the realm of MSSM scenario with
electroweak (EW) supersymmetric (SUSY) particle masses around
a few hunderd~GeV. However, in view of the stringent
constraints on EW SUSY particles  from the 
direct searches at the LHC, it is essential to perform a comprehensive 
analysis including both the \gmin2\ result as well as the latest 
constraints on neutralinos, charginos and sleptons from the LHC.
On the other hand, the R-parity conserving MSSM naturally predicts a
suitable Dark Matter (DM) 
candidate in terms of the lightest neutralino as the lightest
SUSY particle (LSP)~\cite{Go1983,ElHaNaOlSr1984}. Therefore, it seems
only natural to include the experimental constraint on the DM density
and limits from direct detection in the analysis to further constrain
the parameter space of our interest. 

In these proceedings, following \citeres{CHS1,CHS2,CHS3}, we review
the mass ranges of EW superpartners that can successfully explain the \gmin2\
anomaly while being in agreement with all the relevant
experimental data. In \citeres{CHS1,CHS2} we employed the constraint
coming from \refeq{gmtdiff-BNL} to constraint the EW MSSM parameter space.
An updated analysis using the latest experimental world average
can be found in \citere{CHS3}, where we show that the new \gmin2\ result
(\refeq{gmtdiff-new}) confirms the predictions made in \citere{CHS1}
concerning the upper limits on the masses of the 
(next-to-) lightest SUSY particles. We include the latest LHC searches 
via recasting in \CM~\cite{Drees:2013wra,Kim:2015wza, Dercks:2016npn}.
For the DM, we consider two main scenarios depending on whether the LSP
can account for the full relic density or it can be only a subdominant
component of the total DM content of the universe. For the former
scenario, we consider a mixed bino/wino LSP while the latter opens up
the possibility of wino and higgsino DM.
We observe that the combined data helps to narrow down
the allowed parameter region, providing clear targets for possible future
$e^+e^-$~colliders, such as the ILC~\cite{ILC-TDR,LCreport} or
CLIC~\cite{CLIC,LCreport}.


\section{The EW sector of MSSM}
\label{sec:model-constraints}

We give a very brief description of the EW sector of MSSM, consisting of
charginos, neutralinos and sleptons. 
The masses and mixings of the charginos and neutralinos are determined
by $U(1)_Y$ and $SU(2)_L$ gaugino masses $M_1$ and $M_2$, the Higgs
mixing parameter $\mu$ and the ratio of the two
vacuum expectation values (vevs) of the two Higgs doublets of MSSM,
$\tb = v_2/v_1$.
This results in four neutralinos and two charginos 
with the mass ordering $\mneu1 < \mneu2 < \mneu3 <\mneu4$
and $\mcha1 < \mcha2$. Considering the size and sign of the anomaly,
it is sufficient for
our analysis to focus on positive values of $M_1$, $M_2$ and $\mu$
~\cite{CHS1}. 
For the sleptons, we choose common soft
SUSY-breaking parameters for all three generations, $\mL$ and $\mR$. 
We take the trilinear coupling
$A_l$ ($l = e, \mu, \tau$) to be zero for all the three generations of
leptons. 
In general we follow the convention that $\Sl_1$ ($\Sl_2$) has the
large ``left-handed'' (``right-handed'') component.
The symbols equal for all three generations, $\msl1$ and $\msl2$, but we
also refer to scalar muons directly, $\msmu1$ and $\msmu2$.

Following the stronger experimental limits from the
LHC~\cite{ATLAS-SUSY,CMS-SUSY},
we assume that the colored sector of the MSSM is sufficiently heavier
than the EW sector, and does not play a role in this analysis. For the
Higgs-boson sector we assume that the radiative corrections
originating largely from the top/stop sector brings the light
$\cp$-even Higgs boson mass in the experimentally observed region,
$\Mh \sim 125 \gev$. This naturally yields stop masses in the TeV
range~\cite{Bagnaschi:2017tru,Slavich:2020zjv}, in agreement 
with the above assumption. We have not considered
$\CP$-violation in this study, i.e.\ all parameters are real.
$\MA$ has also been set to be above the TeV scale. Consequently, we
do not include explicitly the possibility of $A$-pole annihilation,
with $\MA \sim 2 \mneu1$.
Similarly, we do not consider $h$-~or $Z$-pole annihilation (see,
e.g., \citere{Carena:2018nlf}), as such a 
light neutralino sector likely overshoots the \gmin2\
contribution~\cite{CHS1}.


\section {Relevant constraints}
\label{sec:constraints}

The most important constraint that we consider comes from the \gmin2\
result. We use \refeq{gmtdiff-new} (and in some older results
also \refeq{gmtdiff-BNL}) as a cut at the $\pm2\,\sig$ level. 
However, it is worth mentioning here that we did not take into account 
the results of  the new lattice calculation for the leading order hadronic
vacuuum polarization (LO HVP) contribution~\cite{Borsanyi:2020mff}, 
which have also not been used in the new theory world
average, \refeq{gmt-sm}~\cite{Aoyama:2020ynm}, but would certainly lead to
significant change in our conclusion if turns out to be true, see also
the discussions
in~\citeres{Lehner:2020crt,Crivellin:2020zul,Keshavarzi:2020bfy}. 

We remind that 
the main contribution to \gmin2\ in MSSM at the one-loop level comes from
diagrams involving $\cha1-\Sn$ and $\neu1-\tilde \mu$ loops. 
In our analysis the MSSM contribution to \gmin2\
at two loop order is calculated using {\tt GM2Calc}~\cite{Athron:2015rva},
implementing two-loop corrections
from \cite{vonWeitershausen:2010zr,Fargnoli:2013zia,Bach:2015doa}
(see also \cite{Heinemeyer:2003dq,Heinemeyer:2004yq}).

Various other constraints that are taken into account
comprises the following:
\begin{itemize}
\item{\bf Vacuum stability constraints:}
All points are checked to possess a stable and correct EW vacuum, e.g.\
avoiding charge and color breaking minima, using
the public code {\tt Evade}~\cite{Hollik:2018wrr,Robens:2019kga}.

\item {\bf  Constraints from the LHC:}
All relevant EW SUSY searches are taken into account, mostly via
\CM~\cite{Drees:2013wra,Kim:2015wza, Dercks:2016npn}, where many
analyses had to be implemented newly~\cite{CHS1}.
We also take into account the latest constraints from the disappearing track
searches at the LHC~\cite{Aaboud:2017mpt,Sirunyan:2020pjd}.
These become particularly important for wino DM scenario
where the mass gap between $\chapm1$ and $\neu1$ can be $\sim$ a few
hundred $\mev$.

\item
{\bf Dark matter relic density and direct detection constraints:}
We use the latest result from Planck~\cite{Planck}.
\begin{align}
\Omega_{\rm CDM} h^2 \; = \; 0.120 \pm 0.001 \, .
\label{OmegaCDM}
\end{align}
For the wino and higgsino DM cases, we take the relic density as an {\it
upper} limit 
(evaluated from the central value plus $2\,\sig$).
The relic density in
the MSSM is evaluated with
\MO~\cite{Belanger:2001fz,Belanger:2006is,Belanger:2007zz,Belanger:2013oya}.

We employ the constraint on the spin-independent
DM scattering cross-section $\ssi$ from
XENON1T~\cite{XENON}, evaluating the theoretical prediction using \MO.
For parameter points with $\Omega_{\tilde \chi} h^2 \; \le \; 0.118$
($2\,\sig$ lower limit from Planck~\cite{Planck}), we scale the cross-section
with a factor of ($\Omega_{\tilde \chi} h^2$/0.118)
to account for the fact
that $\neu1$ provides only a fraction of the total DM relic density of
the universe.

\end{itemize}


\section{Parameter scan}
\label{sec:scan}
\smallskip
We scan the relevant MSSM parameter space to obtain lower and {\it upper}
limits on the lightest neutralino, chargino and slepton masses.
The three scan regions  that cover the  complete 
parameter space under consideration are as given below:
\begin{enumerate}[{\bf (A)}]
\item {\bf  bino/wino DM with $\chapm1$-coannihilation:}
\begin{eqnarray}
&100 \gev \leq M_1 \leq 1 \tev \;,
  \quad M_1 \leq M_2 \leq 1.1 M_1\;, \nonumber \\
  &1.1 M_1 \leq \mu \leq 10 M_1, \;
  \quad 5 \leq \tb \leq 60, \; \notag\\
  &\quad 100 \gev \leq \mL \leq 1 \tev, \; \quad \mR = \mL~.
\label{cha-coann}
\end{eqnarray}
\item {\bf Higgsino DM:}
\begin{align}
  100 \gev \leq \mu \leq 1.2 \tev \;,
  \quad 1.1 \mu \leq M_1 \leq 10 \mu\;, \notag \\
  \quad 1.1  \mu \leq M_2 \leq 10 \mu, \;
  \quad 5 \leq \tb \leq 60, \; \notag\\
  \quad 100 \gev \leq \mL, \mR \leq 2  \tev~.
\label{wino-dm}
\end{align}
\item{\bf Wino DM}
\begin{align}
  100 \gev \leq M_2 \leq 1.5 \tev \;,
  \quad 1.1 M_2 \leq M_1 \leq 10 M_2\;, \notag \\
  \quad 1.1 M_2 \leq \mu \leq 10 M_2, \;
  \quad 5 \leq \tb \leq 60, \; \notag\\
  \quad 100 \gev \leq \mL, \mR \leq 2 \tev~.
\label{wino-dm}
\end{align}

\end{enumerate}

In all the scans we choose flat priors of the parameter space and
generate \order{10^7} points.
We use {\tt SuSpect}~\cite{Djouadi:2002ze}
as spectrum and SLHA file generator. The points are required
to satisfy the $\chapm1$ mass limit from LEP~\cite{lepsusy}. 
The SLHA output files
from {\tt SuSpect} are then passed as input to {\tt GM2Calc} and \MO~for
the calculation of \gmin2 and the DM observables, respectively. The parameter
points that satisfy the \gmin2 and DM constraint, 
and additionally 
the vacuum stability constraints checked with {\tt Evade} are
then passed to the final step to be checked against the latest LHC
constraints implemented in \CM. The branching ratios of the
relevant SUSY particles are computed using
{\tt SDECAY}~\cite{Muhlleitner:2003vg} and given as input to \CM.


\section{Results}
\label{sec:results}
In this section we review some of the results for the 
scenarios defined above~\cite{CHS1,CHS2,CHS3}. 
We follow the analysis flow as described above
and denote the points surviving certain constraints
with different colors:
\begin{itemize}
\item grey (round): all scan points.
\item green (round): all points that are in agreement with \gmin2.
\item blue (triangle): points that additionally give the correct relic density.
\item cyan (diamond): points that additionally pass the DD constraints.
\item red (star):  points that additionally pass the LHC constraints.
\end{itemize}

\begin{figure}[htb!]
\begin{subfigure}[b]{0.48\linewidth}
        \centering\includegraphics[width=\textwidth]{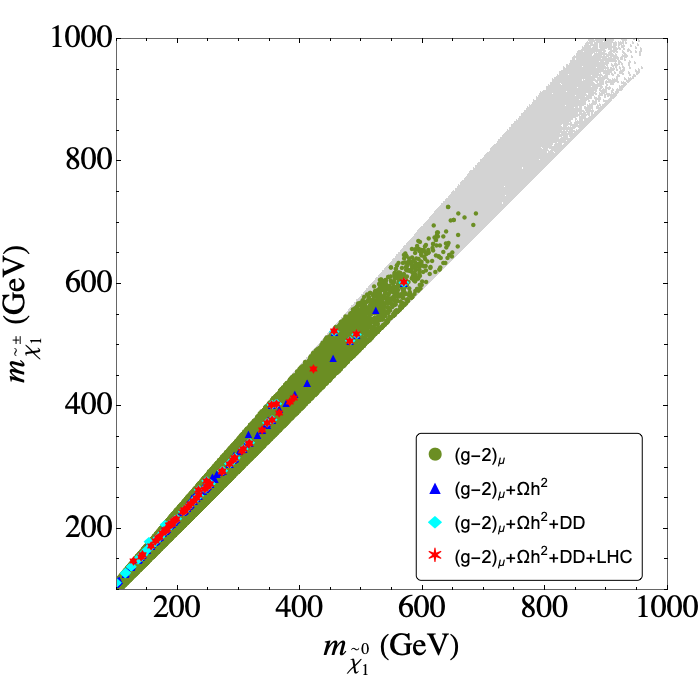}
        \caption{}
        \label{}
\end{subfigure}
~
\begin{subfigure}[b]{0.48\linewidth}
        \centering\includegraphics[width=\textwidth]{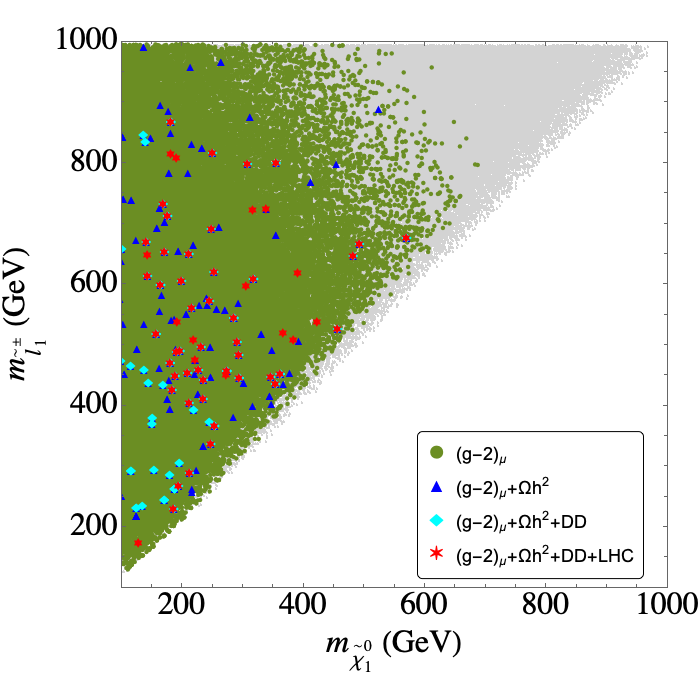}
        \caption{}
        \label{}
\end{subfigure}
\caption{The results of our parameter scan for  the bino/wino
        $\cha1$-coannihilation scenario in the $\mneu1$--$\mcha1$ plane
        (left) and $\mneu1$--$\msl1$ plane (right). 
For the color coding: see text.
}
\label{fig:charco}
\end{figure}

In \reffi{fig:charco} we show our results for the bino/wino
$\cha1$-coannihilation scenario  in the $\mneu1$-$\mcha1$ (left)
and $\mneu1$-$\msl1$ (right) planes~\cite{CHS3}. 
Starting with the \gmin2\ constraint (\refeq{gmtdiff-new}) (green points)
in the $\mneu1$--$\mcha1$ plane, one can
observe a clear upper limit of about $700 \gev$.
Applying the CDM constraints reduce the upper limit further.
The LHC constraints, corresponding to the ``surviving'' red points
(stars), do not yield a further reduction from above, but cut (as
anticipated) only points in the lower mass region.
The LHC constraint which is effective in this parameter plane is the
one designed for compressed spectra~\cite{Aad:2019qnd}.
Other LHC constraint that is effective in this case is the
bound from slepton pair production leading to dilepton and $\met$
in the final state~\cite{Aad:2019vnb}.
Thus, the experimental data set an upper as well as a lower bound,
yielding a clear search target for the upcoming LHC runs, and in
particular for future $e^+e^-$ colliders.

The distribution of the lighter slepton mass (where it should be kept in
mind that we have chosen the same masses for all three generations),
as in the $\mneu1$-$\msl1$ plane, is shown
in the right plot of \reffi{fig:charco}.
The \gmin2\ constraint is satisfied in a triangular region with its tip
around $(\mneu1, \msl1) \sim (\htrs{700} \gev, \htrs{800} \gev)$.
This is slightly reduced when the DM constraints are
taken into account.
The LHC constraints cut out lower slepton masses, going up to
$\msl1 \lsim 400 \gev$, as well as part of the very low $\mneu1$
points nearly independent of $\msl1$. Details on these cuts can be found
in \citere{CHS1}.

\begin{figure}[htb!]
\centering
\begin{subfigure}[b]{0.48\linewidth}
        \centering\includegraphics[width=\textwidth]{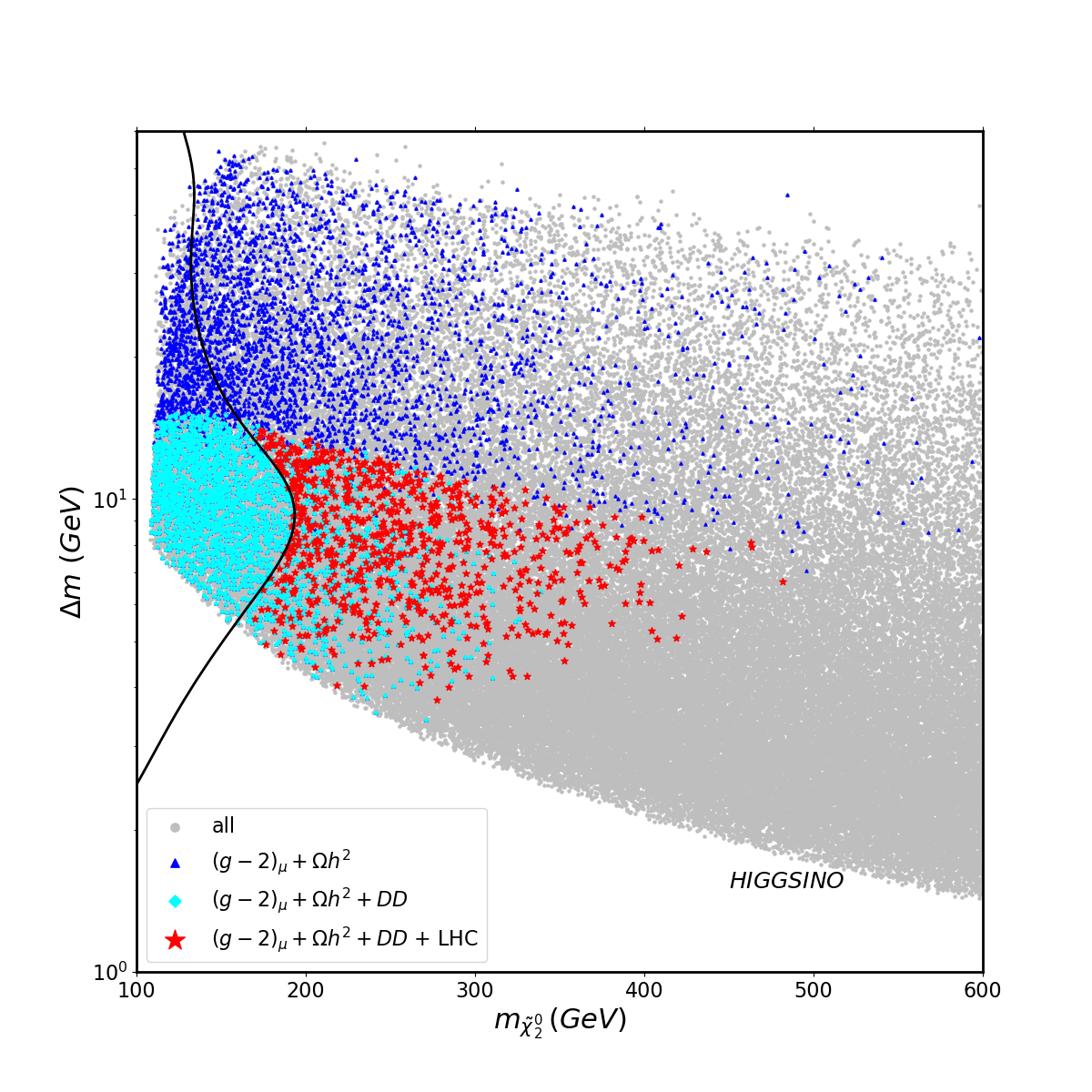}
        \caption{}
        \label{}
\end{subfigure}
~
\begin{subfigure}[b]{0.48\linewidth}
        \centering\includegraphics[width=0.87\textwidth]{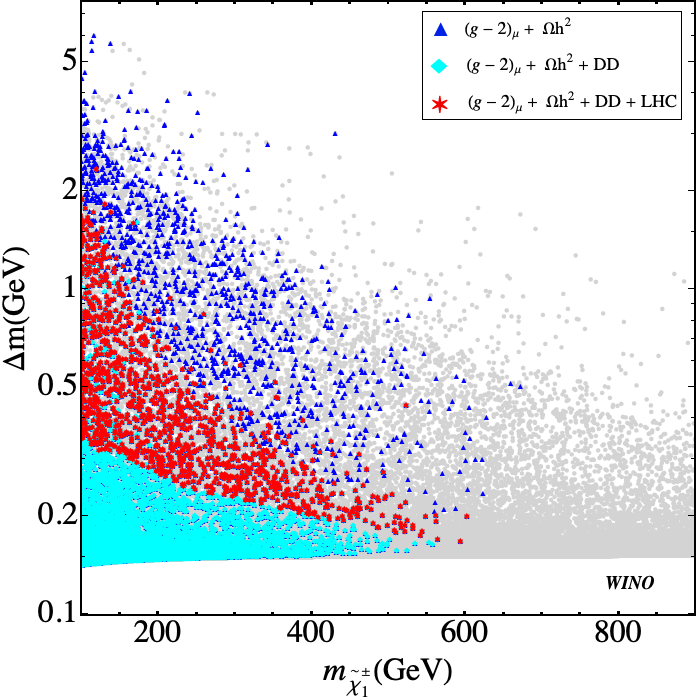}
        \caption{}
        \label{}
\end{subfigure}
\caption{The results of our parameter scan in the
$\mneu2$--$\Delta m(= \mneu2-\mneu1)$ plane for the higgsino DM scenario
(left) and $\mcha1$--$\Delta m(= \mcha1-\mneu1)$ plane 
for the wino DM scenario (right)
(\gmin2\ limits from \refeq{gmtdiff-BNL}).
For the color coding: see text.
}
\label{hiwi}
\end{figure}

Our results for the higgsino and wino DM scenarios are presented
in \reffi{hiwi}~\cite{CHS2}. We show our results in the 
$\mneu2$--$\Delta m(= \mneu2-\mneu1)$ plane 
for the higgsino DM scenario (left) and in the
$\mcha1$--$\Delta m(= \mcha1-\mneu1)$ plane
for the wino DM scenario (right), where the \gmin2 limit here
corresponds to \refeq{gmtdiff-BNL}.
No green points are visible in these plots as all the points that pass
the \gmin2\ constraint are also in agreement with the DM relic density
constraint, resulting in only blue points. For the higgsino DM case, we
also explicitly show the constraint from compressed spectra
searches~\cite{Aad:2019qnd} 
as a black line\footnote{This is an updated plot compared to \citere{CHS2}
due to a small correction in the determination of the compressed search
limits.}. 
In the case of wino DM the relevant LHC constraint are the disappearing
track searches~\cite{Aaboud:2017mpt,Sirunyan:2020pjd},
due to the relatively long life-time of the NLSP, the
light chargino.
In both scenarios the combination of \gmin2, DM limits and LHC searches
put an upper limit on the (N)LSP masses. They are found at
$\sim 500 (600) \gev$ for higgsino (wino) DM. As for the case of
bino/wino DM, clear search targets are set for future LHC runs, and in
particular for the ILC and CLIC.
For a more detailed description of these two scenarios see \citere{CHS2}.

\section{Future linear collider prospects}
\label{sec:future}

In this section we briefly discuss the prospects for the direct detection of
the (relatively light) EW particles at 
possible future $e^+e^-$ colliders such as
ILC~\cite{ILC-TDR,LCreport} or CLIC~\cite{CLIC,LCreport},
which can reach energies up to $1 \tev$ and $3 \tev$, respectively.
We evaluate the cross-sections for various SUSY pair
production modes for the energies currently foreseen in the run plans
of the two colliders.  The anticipated energies and integrated
luminosities are listed in \refta{tab:ee-sqrtS}. The cross-section
predictions are based on tree-level results, obtained as
in~\cite{Heinemeyer:2017izw,Heinemeyer:2018szj}, where it was shown
that the full one-loop corrections can amount up to 10-20\%\,%
\footnote{Evaluation of the full one-loop corrections as 
in \cite{Heinemeyer:2017izw,Heinemeyer:2018szj} would require the
determination of the preferred renormalization scheme for each
point individually (see \cite{Fritzsche:2013fta} for details).}%
. We do not attempt any rigorous experimental analysis,
but follow the idea that to a good approximation,
final states with the sum of the masses smaller than the
center-of-mass energy can be
detected~\cite{Berggren:2013vna,PardodeVera:2020zlr,Berggren:2020tle}. 
We also note that in case of several EW SUSY particles in reach of an
$e^+e^-$ collider, large parts of the overall SUSY spectrum can be
measured and fitted~\cite{Baer:2019gvu}.

\begin{table}[!htb]
\begin{center}
\renewcommand{\arraystretch}{1.4}
\begin{tabular}{|c|c|c||c|c|c|}
\hline
Collider & $\sqrt{s}$ [GeV] & $\cL_{\rm int}$ $[\iab]$ & 
Collider & $\sqrt{s}$ [GeV] & $\cL_{\rm int}$ $[\iab]$ \\
\hline
ILC & 250 & 2 & CLIC & 380 & 1 \\
    & 350 & 0.2 &    & 1500 & 2.5 \\
    & 500 & 4 &      & 3000 & 5 \\
    & 1000 & 8 &     &      & \\
\hline
\end{tabular}
\caption{Anticipated center-of-mass energies, $\sqrt{s}$ and
    corresponding integrated luminosities, $\cL_{\rm int}$ at
    ILC~\cite{Barklow:2015tja,Fujii:2017vwa} and
    CLIC~\cite{Robson:2018zje} (as used in \cite{deBlas:2019rxi}).}
\label{tab:ee-sqrtS}
\renewcommand{\arraystretch}{1.2}
\end{center}
\end{table}

\begin{figure}[htb!]
  \centering
  \begin{subfigure}[b]{0.48\linewidth}
    \centering\includegraphics[width=\textwidth]{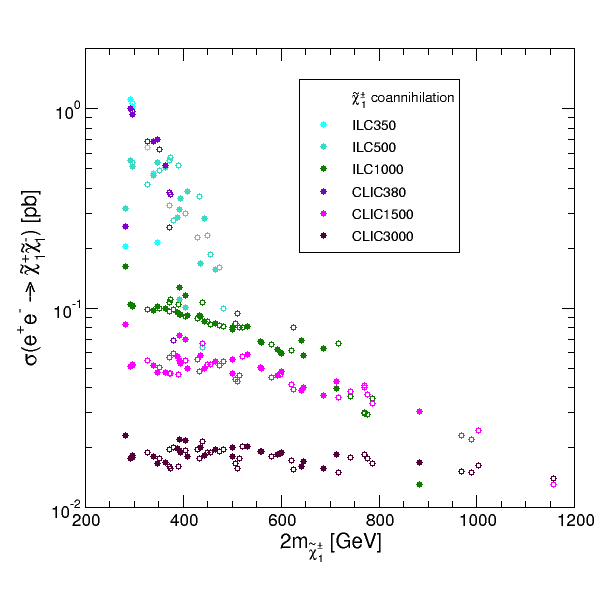}
    \caption{}
    \label{}
  \end{subfigure}
  ~
  \begin{subfigure}[b]{0.48\linewidth}
    \centering\includegraphics[width=\textwidth]{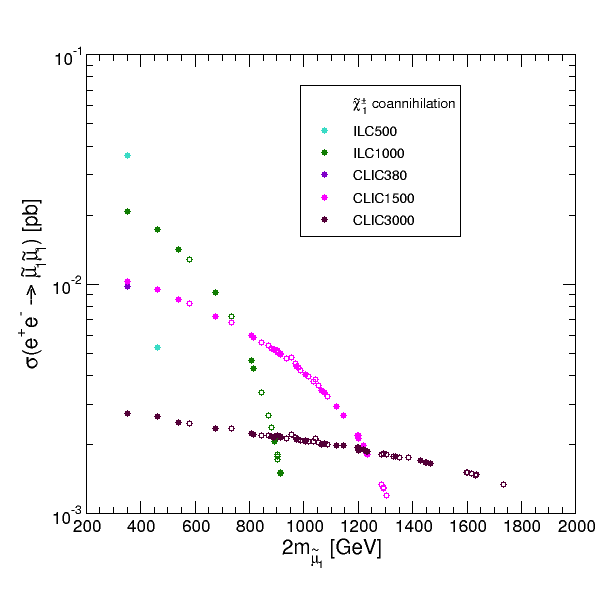}
    \caption{}
    \label{}
  \end{subfigure}
  \caption{cross-section predictions for
  $e^+e^- \to \chapm1\champ1$ (left) and
  $e^+e^- \to \Smu1\Smu1$ (right) for the bino/wino $\cha1$-coannihilation case at
  the ILC and CLIC as a function of the sum of the final state masses.
  Open (filled) circles: see text.}
\label{ee-char}
\end{figure}

In \reffi{ee-char} we show the cross-section predictions for
$e^+e^- \to \chapm1\champ1$ (left) and
$e^+e^- \to \Smu1\Smu1$ (right) in the bino/wino $\cha1$-coannihilation
case as a function of the sum of the final state masses~\cite{CHS1}. 
The points shown in different shades of
green (violet) indicate the cross-sections at the various ILC (CLIC)
energies. All shown points (open and filled) are in agreement with the
old \gmin2\ result, see \refeq{gmtdiff-BNL} (Filled circles indicate a hypothetical future measurement as
discussed in \citere{CHS1}). Using the updated
result, \refeq{gmtdiff-new} would not change this picture in a relevant
way. 
The upper limits on $\mneu1$ of about $570 \gev$ 
for the old (and new) \gmin2\ constraint implies that with
$\sqrt{s} = 1000 \gev$ a considerable part of the allowed region can be
covered.
The reach could become even stronger in the case of the future
\gmin2\ constraint. With the same central value but only slightly better
precision the upper limits on $\mneu1$ go down to $\sim 450 \gev$, 
implying effectively a full coverage at a $1000 \gev$ collider.
In case of smuon pair production, as shown in the right plot
of \reffi{ee-char}, energies up to $\sim 1800 \gev$ would be needed to
fully cover the allowed parameter space.

All obtained cross-section predictions
for the kinematically accessible parameter points are above $10^{-2}$~pb
for chargino production and above $10^{-3}$~pb for smuon pair
production. For each \iab\ of integrated luminosity this corresponds to
10000 (1000) events for chargino (smuon) pair production, which should
make these particles easily accessible, see \refta{tab:ee-sqrtS}, if
they are in the kinematic reach of the collider.

\smallskip
The above shown example cross-sections clearly show that
at least some particles are guaranteed
to be discovered at the higher-energy stages of the ILC and/or CLIC. If
the upcoming runs from the MUON G-2 experiment further confirm the
deviation of $\amu^{\rm exp}$ from the SM prediction, the case for
future $e^+e^-$ colliders is clearly strengthened.

\begin{figure}[htb!]
\centering
  \includegraphics[width=0.6\textwidth]{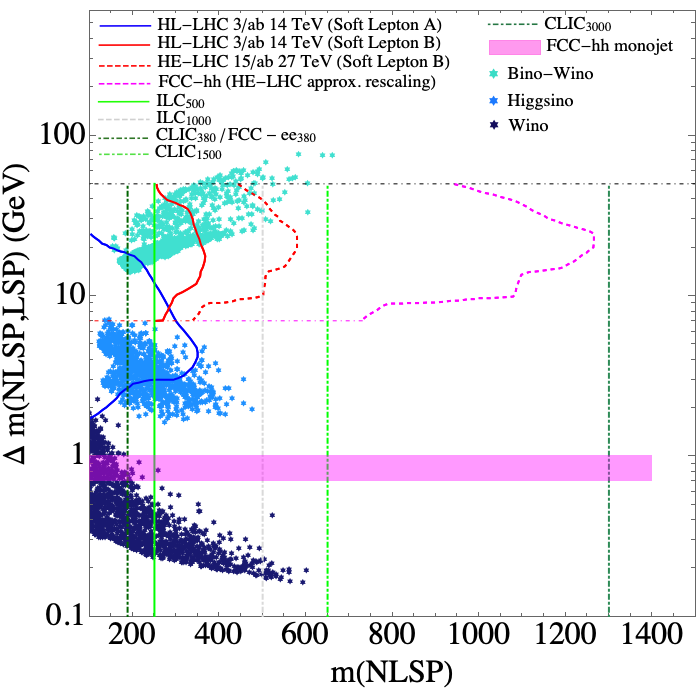}
  \caption{$\mcha{1}$--$\De m$ plane
  with anticipated limits from compressed spectra searches at various
  future colliders, 
  taken from \protect\citere{Strategy:2019vxc}. Disappearing
  track searches are not included.
  Shown in blue, dark blue, turquoise are the points surviving all current
  constraints in the case of higgsino DM, wino DM and bino/wino DM with
  $\cha1$-coannihilation, respectively, with relic density taken only as
  an upper limit.
}
\label{fig:future}
\end{figure}

\smallskip
In \reffi{fig:future} we review the prospects at various high energy colliders
for the compressed spectra searches,
relevant for higgsino DM, wino DM and
bino/wino DM with $\cha{1}$-coannihilation~\cite{CHS2}. 
We show our results in the $\mcha1$--$\De m$ plane
(with $\De m := \mcha1 - \mneu1$), which was presented (upto $\De m = 0.7 \gev$)
in \citere{Strategy:2019vxc} for the higgsino DM case, but also directly
applicable for the wino DM case~\cite{Berggren:2020tle}. In addition to the
anticipated limits from HL-LHC, HE-LHC and FCC-hh, we show
the following projected limits from various high energy linear colliders
sensitive up to the kinematic limit~\cite{Strategy:2019vxc}, looking at
$\cha1\champ1$ or $\neu2\neu1$ production (see
also \citere{Berggren:2020tle} and references therein)

\begin{itemize}
\item ILC with $0.5\,\iab$ at $\sqrt{s} = 500 \gev$ (ILC500): solid
light green.

\item ILC with $1\,\iab$ at $\sqrt{s} = 1000 \gev$ (ILC1000):
gray dashed.

\item CLIC with $1\,\iab$ at $\sqrt{s} = 380 \gev$ (CLIC380):
very dark green dot-dashed.

\item CLIC with $2.5\,\iab$ at $\sqrt{s} = 1500 \gev$ (CLIC1500):
green dot-dashed.

\item CLIC with $5\,\iab$ at $\sqrt{s} = 3000 \gev$ (CLIC3000):
dark green dot-dashed.

\end{itemize}

It can be observed that for the higgsino case, the HL-LHC can cover a 
part of the allowed parameter space, but an exhaustive coverage of the
allowed parameter 
space can be reached only at a high-energy $e^+e^-$ collider with
$\sqrt{s} \lsim 1000 \gev$ (i.e.\ ILC1000 or CLIC1500). For the wino DM,
the $\De m$ is so small that
it largely escapes the HL-LHC searches (but may partially be detectable
at the FCC-hh with monojet searches). As in the
higgsino DM case, also here a high-energy $e^+e^-$ collider will be
necessary to cover the full allowed parameter space. While the
currently allowed points would be covered by CLIC1500, a parameter space
reduced further by e.g.\ the improved HL-LHC disappearing track searches,
could be covered by the ILC1000.
The bino/wino parameter points (turquoise) represent a more complicated case
for the future collider analysis, since the limits 
assume a small mass difference between $\neu1$ and $\cha1$ as well as
$pp$ production cross sections for the higgsino case. For bino/wino DM
case, these typically have larger production cross sections (so is the pure
wino), i.e.\ application of these limits to the wino/bino points
in \reffi{fig:future} serves as conservative estimate for $pp$ based limits.
Consequently, it is expected that the HE-LHC or the FCC-hh would cover this
scenario entirely. On the other hand, the $e^+e^-$ limits should be
directly applicable, and large parts of the parameter space will be effectively
covered by the ILC1000, and the entire parameter space by CLIC1500.

\medskip
\noindent
Although we do not consider possibility of $Z$~or $h$~pole annihilation,
it should be noted that in this context an LSP with
$M \sim \mneu1 \sim \MZ/2$ or $\sim \Mh/2$ (with $M = M_1$ or $M_2$ or $\mu$)
would yield a detectable cross-section $e^+e^- \to \neu1\neu1\ga$
in any future high-energy $e^+e^-$ collider. Furthermore, in the case of
higgsino or wino DM,  
this scenario automatically yields other clearly detectable EW-SUSY
signals at future $e^+e^-$ colliders. For bino/wino DM
this would depend on the values of $M_2$ and/or $\mu$.


\section*{Acknowledgments}
I.S.\ thanks S.~Matsumoto for the cluster facility.
The work of I.S.\ is supported by World Premier
International Research Center Initiative (WPI), MEXT, Japan.
The work of S.H.\ is supported in part by the
MEINCOP Spain under contract PID2019-110058GB-C21 and in part by
the AEI through the grant IFT Centro de Excelencia Severo Ochoa SEV-2016-0597.
The work of M.C.\ is supported by the project AstroCeNT:
Particle Astrophysics Science and Technology Centre,  carried out within
the International Research Agendas programme of
the Foundation for Polish Science financed by the
European Union under the European Regional Development Fund.


\newcommand\jnl[1]{\textit{\frenchspacing #1}}
\newcommand\vol[1]{\textbf{#1}}

\newpage{\pagestyle{empty}\cleardoublepage}


\end{document}

%% file: paperdef.tex
\newcommand{\lsim}
{\;\raisebox{-.3em}{$\stackrel{\displaystyle <}{\sim}$}\;}

\newcommand\tb{\tan\beta}

\newcommand\ReDiag{\mathop{%
  \raise .5pt\hbox{[}%
  \widetilde{\mathrm{Re}}%
  \raise .5pt\hbox{]}}}
\newcommand\ReOffDiag{\mathop{%
  \raise .5pt\hbox{$\llbracket$}%
  \widetilde{\mathrm{Re}}%
  \raise .5pt\hbox{$\rrbracket$}}}

\newcommand\cL{{\cal L}}

\newcommand\MZ{M_Z}
\newcommand\Mh{M_h}

\newcommand\MA{M_A}

\newcommand\Sn{\tilde\nu}
\newcommand\Sl{\tilde l}

\newcommand\Smu[1]{\tilde \mu_{#1}}

\newcommand\msl[1]{m_{\Sl_{#1}}}

\newcommand\mL{m_{\tilde l_L}}
\newcommand\mR{m_{\tilde l_R}}

\newcommand\ino[1]{\tilde\chi_{#1}}

\newcommand\chapm[1]{\ino{#1}^\pm}
\newcommand\champ[1]{\ino{#1}^\mp}

\newcommand\cha{\chapm}
\newcommand\mcha[1]{m_{\chapm{#1}}}

\newcommand\neu[1]{\ino{#1}^0}
\newcommand\mneu[1]{m_{\neu{#1}}}

\newcommand\refeq[1]{Eq.~(\ref{#1})}

\newcommand\refta[1]{Tab.~\ref{#1}}

\newcommand\citere[1]{Ref.~\cite{#1}}
\newcommand\citeres[1]{Refs.~\cite{#1}}

\newcommand{\CP}{{\cal CP}}
\newcommand{\cp}{{\CP}}

\newcommand{\tev}{\,\, \mathrm{TeV}}
\newcommand{\gev}{\,\, \mathrm{GeV}}
\newcommand{\mev}{\,\, \mathrm{MeV}}

\newcommand\MO{\texttt{MicrOMEGAs}}
\newcommand\CM{\texttt{CheckMATE}}

\newcommand\ab{\ensuremath{\mbox{ab}}}

\newcommand\iab{\ensuremath{\ab^{-1}}}

\newcommand\msmu[1]{m_{\tilde{\mu}_{#1}}}

\newcommand{\sig}{\sigma}

\def\order#1{\ensuremath{{\cal O}(#1)}}
\def\reffi#1{\mbox{Fig.~\ref{#1}}}

\def\ga{\gamma}
\def\De{\Delta}

\def\gmin2{\ensuremath{(g-2)_\mu}}
\def\amu{\ensuremath{a_\mu}}
\def \met  {\mbox{${E\!\!\!\!/_T}$}}
\newcommand{\ssi}{\ensuremath{\sig_p^{\rm SI}}}

\definecolor{Orange}{named}{orange}
\definecolor{Purple}{named}{purple}
\definecolor{Lightblue}{cmyk}{0.9,0.1,0.1,0.3}
\definecolor{dgelborange}{cmyk}{0.,0.3,0.5, 0.}
\definecolor{Lila}{rgb}{0.5,0.,1}
\definecolor{Darkgreen}{rgb}{0.,.7,0.2}